\documentclass[useAMS,usenatbib]{mn2e}
\usepackage{amsmath}
\usepackage{amsfonts}
\usepackage{amssymb,graphicx}
\usepackage{lscape}

\title[The unusual radio transient in M\,82: an SS\,433 analogue?]{The unusual radio transient in M\,82: an SS\,433 analogue?}
\author[T. D. Joseph, T. J. Maccarone and R. P. Fender]{T. D. Joseph$^{1}$ \thanks{E-mail: tdj1f08@soton.ac.uk; tjm@phys.soton.ac.uk; r.fender@soton.ac.uk}, T. J. Maccarone$^{1}$ \footnotemark[1] and R. P. Fender$^{1}$  \footnotemark[1]\\
$^{1}$University of Southampton, Southampton, SO17 1BJ, United Kingdom}
\begin{document}

\date{Accepted ????. Received ????; in original form ????}

\pagerange{\pageref{firstpage}--\pageref{lastpage}} \pubyear{????}

\maketitle

\label{firstpage}

\begin{abstract}
In this paper we discuss the recently discovered radio transient in the nuclear region of M\,82. It has been suggested that this source is an X-ray binary, which, given the radio flux density, would require an X-ray luminosity, $L_{\rm X}\,\sim6\times10^{42}$\,erg\,s$^{-1}$ if it were a stellar mass black hole that followed established
empirical relations for X-ray binaries. The source is not detected in the analysis of the X-ray archival data. Using a 99\,\% confidence level upper limit we find that $L_{\rm X} \leq 1.8 \times 10^{37}$\,erg\,s$^{-1}$ and $1.5 \times 10^{37}$\,erg\,s$^{-1}$, using powerlaw and disk blackbody models respectively. The source is thus unlikely to be a traditional microquasar, but could be a system similar to SS\,433, a Galactic microquasar with a high ratio of radio to X-ray luminosity.

\end{abstract}


\section{Introduction}

\par
M\,82 is prolific star-forming galaxy at a distance of 3.6\,Mpc \citep{1994ApJ...427..628F}. Its proximity therefore makes it an ideal place to study star formation. To this end, M\,82 has been well monitored by various observatories at several wavelengths for decades \citep[e.g.][]{1980ApJ...235..392T, 1981ApJ...246..751K, 1988Natur.334...43B,1994MNRAS.266..455M, 1997A&A...320..378S, 2001A&A...365..571W, 2008MNRAS.391.1384F}. In particular, radio observations have revealed approximately 60 compact radio sources in the central region of M\,82 \citep{2002MNRAS.334..912M}. A quarter of these sources are of unknown origin. In addition to compact sources, frequent radio observations have also found transient sources of an undetermined nature in M\,82 \citep{1985Sci...227...28K, 1994MNRAS.266..455M} and monitored the evolution of radio supernovae \citep[e.g.][]{1994MNRAS.266..455M, 2006MNRAS.369.1221B, 2008MNRAS.391.1384F}.  

\par 
The next generation of radio telescopes should reveal many more such transient sources. These observatories will have the ability to detect a 0.1\,mJy source within minutes. We will then more easily be able to observe transient populations not only in star-forming galaxies like M\,82 where these systems are abundant, but also in galaxies with low star-formation rates like ellipticals. By attempting to determine the nature of the unknown transients such as those in M\,82, we will make the task of studying the large number of transients expected to be observed in the future much easier. 

\par
In this paper we investigate further the most recently discovered radio transient of unknown origin in M\,82 and show that this source is unlikely to be a normal microquasar. We discuss the possibility that the M\,82 source could be an extragalactic analogue of SS\,433, a very unusual Galactic microquasar \citep[][and references therein]{1984ARA&A..22..507M}. Both sources have low X-ray and high radio luminosities. In addition, SS\,433 possesses precessing, helical jets. SS\,433 also has highly blue- and redshifted optical emission lines associated with it \citep{1979ApJ...230L..41M}. If the M\,82 transient is an extragalactic SS\,433 like microquasar, it will be only the second such source, after the microquasar S\,26 in NGC\,7793, to be discovered \citep{2010Natur.466..209P}. 

\section{Properties of the source}
The radio source was first discovered with the Multi-Element Radio-Linked Interferometer Network (MERLIN) early in May 2009 by \cite{2009ATel.2073....1M} as part of a continuous monitoring of the recent supernova SN 2008iz in M\,82. \cite{2009ATel.2078....1B} later also reported a detection of the source on 2009 April 30 with the Very Long Baseline Array (VLBA). 

\par
For the last 15 to 20 years M\,82 has been observed at radio frequencies at intervals of six months to one year. In that time, only two other transient sources of unknown nature have been detected in the galaxy \citep{1985Sci...227...28K, 1994MNRAS.266..455M}. This most recent transient source was not detected by \cite{2009ATel.2073....1M} in the period 2009 April 24th to 27th to a 3\,$\sigma$ upper limit of $< 0.2$\,mJy/\,beam at 4.9\,GHz. \cite{2009ATel.2078....1B} found that the source was not detected in their observations taken on 2009 April 8 (3\,$\sigma$ limit of 0.6\,mJy at 22\,GHz) and 2009 April 27 (3\,$\sigma$ limit of 0.9\,mJy at 43\,GHz and 0.7\,mJy at 22\,GHz). Moreover, \cite{2010MNRAS.404L.109M} report that the source has not been detected in other wavebands previously or concurrently with their radio detections, including infrared \citep{2009ATel.2131....1F}, optical (Matilla, private communication) and X-ray observations \citep{2009ATel.2080....1K}. 

\par
The source is situated in the nuclear region of the galaxy, at a position of RA 09$^{h}$55$^{m}$52$^{s}$.5083, Dec. 69$^{\circ}40'45''.410$ (J2000) with 5\,mas error in each coordinate. The VLBA observations show flux densities of 1.1\,mJy and 0.5\,mJy at 1.6\,GHz and 4.8\,GHz respectively. The observations from the first week of 2009 May revealed a peak radio flux of between 0.6 and 0.7\,mJy at 4.9\,GHz. Follow up observations showed that the flux remained roughly constant over the next 150 days \citep{2010MNRAS.404L.109M}. The radio spectral index, $\alpha$, also remained unchanged at $-$0.7 during this time, where $F_{\nu} \propto \nu^{\alpha}$. MERLIN data taken in 2010 January (over 240 days after the initial detection) show that the source is roughly 15\,mas in size.

\section{X-ray Observations}

The nucleus of M\,82 was observed with Chandra ACIS-S on 2009 April 17 (ObsID 10025) and 2009 April 29 (ObsID 10026). The data files were downloaded from the Chandra on-line archive. The data were checked for background flares and spectra were produced using the CIAO  script \emph{psextract}. We used a circular region of radius 2$''$ centered on the radio source to extract the source and background spectra from ObsID 10026 and 10025 respectively. 

\par
The spectra were analysed using the Interactive Spectral Interpretation System (ISIS), version 1.6.1. As ObsID 10025 was taken well before the source was first seen in radio, we used this observation as the background spectrum for our analysis. Given the initial, sharp radio variability, we expect that the variable component of the X-ray flux should be comparable to the X-ray flux itself. We were unable to use the standard detection experiment approach because of the strong diffuse X-ray emission in the centre of M\,82. The data were grouped into bins of 50 counts to increase signal-to-noise; binning to the standard signal-to-noise ratio of 5 produced too few bins for fitting.  Only data in the range 0.3$-$10\,keV were included in the spectral fitting. Due to the low number of counts only two spectral models were used, namely disk blackbody (diskBB) and power law (PL) \citep[see][]{1989PASJ...41...97M}. We fixed the PL index, $\Gamma = 2$ and the diskBB temperature, $T_{in} = 1$\,keV. The Gehrels statistic was used to calculate errors as it gives a more reliable fit to data with low count rates \citep{1986ApJ...303..336G}. Both the diskBB and PL models give similar results. 

\par
\begin{figure}
 \centering
\includegraphics[width=0.45\textwidth, height=0.5\textwidth]{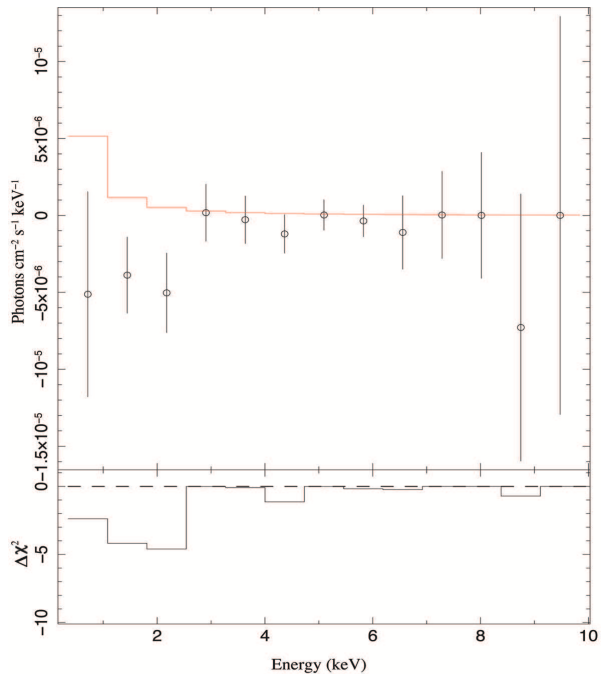}
\caption{The PL model fit (99\,\% confidence level upper limit) of X-ray (0.3$-$10\,keV) data for the M\,82 source. The$\Delta \chi^2$ contributions from each bin are shown in the bottom panel, with the sign indicating when the data are above or below the model.}
\label{PL}
\end{figure}

The spectral fitting of the data yielded no clear detection of significant X-ray flux. In fact, the background spectrum had a higher number of counts than the source spectrum. We calculated the upper limits corresponding to a 99\,\% confidence level for the X-ray luminosity and found that the unabsorbed luminosities are  1.8$\times10^{37}$\,erg\,s$^{-1}$ and 1.5$\times10^{37}$\,erg\,s$^{-1}$ for PL and diskBB respectively, using Galactic column density, $n_H = 4 \times 10^{20}$\,cm$^{-2}$ \citep{1990ARA&A..28..215D}.  The spectral fit for the 99\,\% confidence level limit of the PL model is shown in Fig.\,\ref{PL}. The data are in the range 0.3$-$10\,keV. The bottom panel shows the $\Delta \chi^2$ value from each bin.

\par
The source is situated at a local minimum in foreground column density with a value of 3.5$\times10^{22}$\,cm$^{-2}$ estimated from CO maps \citep{2002ApJ...580L..21W}. Using this measured column density, the upper limits corresponding to a 99\,\% confidence level for $L_{\rm X}$ in the $0.3-10$\,keV range is 4.9$\times10^{37}$\,erg\,s$^{-1}$ (PL) and 3.1$\times10^{37}$\,erg\,s$^{-1}$ (diskBB). The X-ray non-detection is therefore unlikely to be caused by absorption in M\,82. 

\section{Properties of Galactic black hole binaries}
\cite{2010MNRAS.404L.109M} discuss various possibilities regarding the nature of the source, including that the source is an extragalactic microquasar. Other ideas put forward are that the source might be an AGN in the centre of M\,82 or an unusual radio supernova. However, they also note that both these explanations are unlikely. If the source is an AGN it would require the supermassive black hole to be significantly displaced from the dynamical centre of the galaxy. The source is also too faint to be a Type I supernova and the unchanging spectral index argues against young radio supernovae in general. 

\par
In this paper we focus on the microquasar scenario. We use the properties of black hole binaries (BHBs) to predict various parameters, such as X-ray luminosity, for our source.

\par
An unambiguous correlation has been found between X-ray luminosity, $L_{\rm X}$, and radio luminosity, $L_{\rm radio}$, for stellar mass black holes in the hard X-ray state (always the case below ~1\% Eddington), of the form $L_{\rm radio} \propto L_{\rm X}^b$ where $0.6 \leq b \leq 0.7$ \citep{2003A&A...400.1007C, 2003MNRAS.344...60G, 2006MNRAS.370.1351G}. At higher Eddington ratios, X-ray binaries can enter phases of radio flaring associated with state transitions. As discussed in \cite{2004MNRAS.355.1105F} (see for example their Fig.\,2), in such states the radio/X-ray correlation is still broadly consistent with the data, albeit with a larger scatter (of around 1 dex in $L_{\rm radio}$, corresponding to a scatter in $L_{\rm X}$ of $\sim 1.4$ dex). This radio/X-ray correlation has been shown to extend, with the addition of a mass term, $M$, to active galactic nuclei (AGN), in a tight correlation for AGN at low Eddington ratios \citep{2004A&A...414..895F}, and with a similar correlation but broader scatter when including a sample of AGN with no restriction on Eddington ratio or state, and which thus must include flaring sources \citep[][hereafter MHdM03]{Merloni03}. The best fit to the MHdM03 `fundamental plane' is of the form $L_{\rm radio} \propto L_{\rm X}^{0.6} M^{0.8}$, and has a scatter of about 2 dex in $L_{\rm radio}$. The current state of knowledge is, therefore, that for `normal' X-ray binaries and AGN there is a common radio/X-ray correlation which is tighter in hard states but which follows a similar, but less precise, correlation even in flaring states. There are of course additional caveats if the radio observations are very sparse (e.g. whether or not the flare peak was caught), but in the case of the M82 transient the sampling is good.

\par
Assuming that the M82 source is a normal BH X-ray binary transient following the Corbel/Gallo relations, the measured radio luminosity implies $L_{\rm X} \sim 6 \times 10^{42}$ erg s$^{-1}$, far in excess of our established upper limit (see Fig.\,\ref{BHB_plot}). Using the MHdM03 fundamental plane and assuming the M\,82 source is Eddington-limited, then $L_{\rm X} = L_{\rm Edd} = 1.3\times10^{38} (M/M_{\odot})$\,erg\,s$^{-1}$ implies $M = 1.9\times10^3$\,M$_\odot$ and $L_{\rm X} = 2.5\times10^{41}$\,erg\,s$^{-1}$. This inferred luminosity suggests that the source should be very bright in X-rays and that it could be an intermediate mass BH system. For the Eddington limited case $L_{\rm X} \propto M$ and thus $L_{\rm radio} \propto L_{\rm X}^{1.4}$, leading to a scatter to 1.4 dex in $L_{\rm X}$, the same as that estimated for the radio/X-ray correlation. This scatter introduces an uncertainty of a factor of about 30 into the implied $L_{\rm X}$. If $L_{\rm X}$ is reduced by a factor of 30, the resulting luminosity is still several orders of magnitude higher than the upper limit estimated from the spectral fits.

\par
It should be noted that $L_{\rm radio}/L_{\rm X}$ could sometimes be higher for HMXBs than for LMXBs. HMXBs have dense stellar winds that might interact with the jets from the BH \citep[see e.g.][]{2003A&A...410L...1R}. Any jet-wind interaction could give rise to synchrotron radiation thereby increasing the radio emission of the system. The mass transfer rate, $\dot{M}$, of the binary system also increases with the mass of the donor star (see \S 5). Additionally, HMXBs in Roche lobe overflow may have donor stars sufficiently massive to allow unstable mass transfer. If this happens, the accretion rates can become highly super-Eddington. When $\dot{M}$ is strongly super-Eddington, the accretion disk is expected to become geometrically thick and the X-rays from the BH are scattered and absorbed \citep[e.g.][]{2006MNRAS.370..399B}. Both the low radiative efficiency in the X-rays and the high radiative efficiency of the jets in radio then serve to increase $L_{\rm radio}/L_{\rm X}$.  

\par
\begin{figure}
\centering
\includegraphics[width=0.5\textwidth]{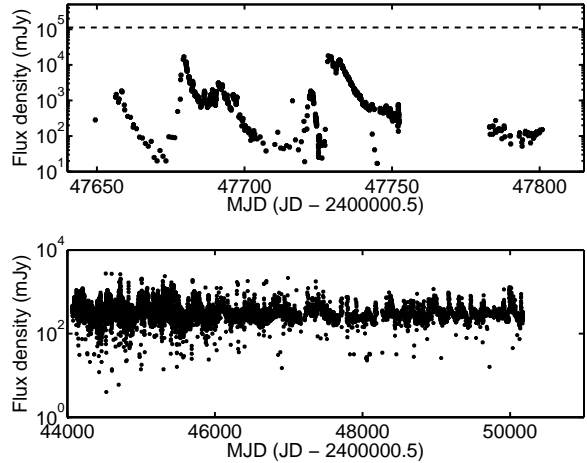}
\caption{Top panel: A 150 day light curve of Cyg X-3 taken with the GBI at 2.25\,GHz \citep{1995AJ....110..290W}. The horizontal line indicates the flux density of our source scaled to the distance of Cyg X-3. Bottom panel: A light curve of SS\,433 taken with the GBI at 2.7\,GHz \citep[see][for data up to December 1985 and procedure for subsequent data downloaded from the GBI website]{1987AJ.....94.1244F}.}
\label{CygX3_SS433}
\end{figure}

Now we compare the M\,82 source to other non-standard BHBs to try and establish whether or not our source could be such a system. Cyg X-3 is the brightest quasi-persistent Galactic BHB at radio wavelengths with a peak flux density of approximately 20\,Jy at 2.25\,GHz. When scaled to the distance of Cyg X-3 \cite[\emph{d}=9\,kpc,][]{2000A&A...357L..25P} the peak flux density of our source is 110\,Jy, nearly ten times that of the brightest Cyg X-3 flare. Thus this source would be exceptionally radio bright if it is a microquasar. Moreover, the spectral index of Cyg X-3 varies greatly during flaring episodes, with $-0.4 \lesssim \alpha \lesssim 1.8$ \citep{1995AJ....110..290W}. It is then unlikely that Cyg X-3 and the source in M\,82 are in the same accretion state. Figure\,\ref{CygX3_SS433} (top panel) shows a radio (2.25\,GHz) light curve of Cyg X-3 for a period of 150 days taken with the Green Bank Interferometer (GBI) \citep{1995AJ....110..290W}. The horizontal line indicates the peak flux density of our source scaled to the distance of Cyg X-3.

\par
Most other Galactic microquasars have radio flaring events that show a rapid increase in flux density over periods of hours to days; the flares then decay back to quiescent levels within a few days to weeks \citep[see e.g.][]{2001ESASP.459..291H,2002MNRAS.331..765B,2002ATel..107....1F,2003MNRAS.342..623S}. The flux density of the M\,82 source increased by at least a factor of five over a period of a few days. Unlike most other microquasars, however, the source then remained at approximately constant flux for several months, with flux density variations of less than a factor of two \citet[see][Fig.2]{2010MNRAS.404L.109M}. 

\par
SS\,433 is an Galactic exotic microquasar that in some respects exhibits radio emission unlike that of other BHBs \citep[e.g.][]{2004ASPRv..12....1F}. A light curve taken with the GBI at 2.7\,GHz for this system is shown in Fig.\,\ref{CygX3_SS433} (bottom panel). The data up to 1985 December were taken from  \cite{1987AJ.....94.1244F}. Subsequent data were downloaded from the GBI website \footnote{http://www.gb.nrao.edu/fgdocs/gbi/gbint.html}; the procedure is the same as in \cite{1987AJ.....94.1244F}. The light curve of SS\,433 is similar to that of our source as its flux density stays roughly constant over a period much greater than the decay time of most microquasar flares. In addition, the spectral index of SS\,433 is $-0.5$ \citep{1998AJ....116.1842D}, similar to that of the M\,82 source.

\par
S\,26 is already thought to be an extragalactic analogue of SS\,433 \citep{2010MNRAS.409..541S}. In contrast to the point-like M\,82 transient, S\,26 has extended structures of radio lobes and X-ray and radio hotspots nearly 300\,pc apart enveloped in a cocoon of gas that has been inflated by the jets. The radio spectral index  varies across the system. In the lobes $-0.7 \lesssim \alpha \lesssim -0.6$, it flattens across the cocoon with $-0.4 \lesssim \alpha \lesssim 0$ and is inverted near the base of the jets where $0 \lesssim \alpha \lesssim 0.4$. The core of S\,26 has an X-ray luminosity of $\sim 7 \times 10^{36}$\,erg\,s$^{-1}$ (0.3 $-$ 10\,keV), but no radio emission has been detected from the core (3\,$\sigma$ upper limit $\sim$ 3$\times 10 ^{33}$\,erg\,s$^{-1}$). This more complex structure suggests that if S\,26 and the M\,82 source are binary systems with similar components (i.e. similar donor star and accretor), then S26 has been undergoing mass transfer for a much longer duration so that it has had a chance to power strong lobes.

\section{Discussion}

\begin{figure}
\centering
\includegraphics[width=0.5\textwidth]{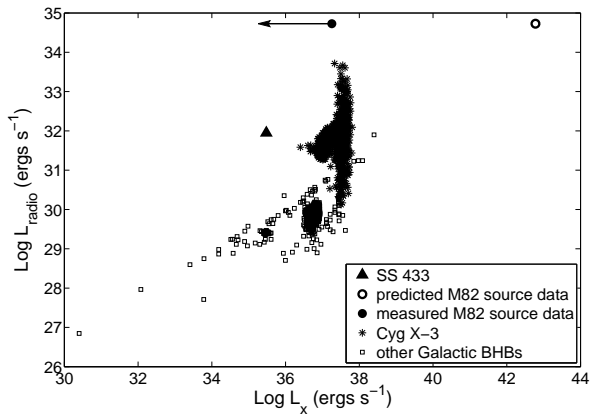}
\caption{Plot showing radio/X-ray correlation for hard state BHBs. The
  squares indicate the BHB data from \citet{2010MNRAS.409..839C}; the
  X-ray data are in the 1$-$10\,keV range and the radio data are
  calculated at 5\,GHz. The triangle shows a representative SS\,433
  data point \citep[2$-$10\,keV and 4.9\,GHz for X-ray and radio data
  respectively;][and references therein]{1984ARA&A..22..507M}. The
  asterisks indicate Cyg X-3 data  \citep[2$-$11\,keV and 4.9\,GHz for
  X-ray and radio data respectively;][]{2003MNRAS.344...60G}. The
 filled and open circles are the actual and estimated (from the
 radio/X-ray correlation) data for the M\,82 source respectively,
 with $L_{\rm X}$ in the 1$-$10\,keV range and $L_{\rm radio}$
 calculated at 4.9\,GHz. The errors have been omitted from the plot
 for clarity; see MHdM03 for error estimates.}
\label{BHB_plot}
\end{figure}

\par
In Fig.\,\ref{BHB_plot} we present the observed upper limit on the X-ray luminosity and the X-ray luminosity predicted from the \cite{2003MNRAS.344...60G} correlation versus the radio luminosity on the plot of the observed X-ray and radio luminosities for BHBs, using the data from \cite{2010MNRAS.409..839C}. For the purpose of this comparison we recalculate $L_{\rm X}$ for the M\,82 source to be in the 1$-$10\,keV range and $L_{\rm radio}$ is calculated from the flux density data at 4.9\,GHz. Data for SS\,433 \citep[and references therein]{1984ARA&A..22..507M} and Cyg X-3 \citep{2003MNRAS.344...60G} are also included in the plot. The X-ray luminosity for the M\,82 source is lower than any other transient Galactic BHB, yet the radio luminosity is exceptionally high. This, along with behaviour of the light curve, makes it unlikely that the source is a normal microquasar. Like the M\,82 source, SS\,433 is also relatively faint in X-rays compared to radio, with $L_{\rm X} \sim 3\times10^{35} $\,erg\,s$^{-1}$ (2$-$10\,keV) and $L_{\rm radio} \sim 7\times 10^{31}$\,erg\,s$^{-1}$ at 4.9\,GHz. The M\,82 transient source could therefore possibly be an extragalactic analogue of this system. 

\par
The fact that the M\,82 source has a significantly higher radio luminosity than SS\,433 could be explained ,perhaps, by a more massive donor star. The mass of the donor star in SS\,433 is thought to be only about 12\,M$_{\odot}$ \citep{2008ApJ...676L..37H}. This mass is at the lower limit for systems powered by thermal timescale accretion onto a BH \citep{2000ApJ...530L..25K}. It is worth bearing in mind that at the distance of M\,82, SS\,433 and the W50 nebula would be difficult to distinguish from a supernova remnant. Additionally, the flux density of W50 ($d=5.5$\,kpc) is 70\,Jy  at 1.4\,GHz \citep{1998AJ....116.1842D}. At the distance of M\,82, the flux density would be 160\,$\mu$Jy spread over an angular scale of 7$''$. At these flux densities such sources will be impossible to detect as they are well below the confusion limit for M\,82 \citep[1.45\,mJy with 1$''$ angular resolution at 4.9\,GHz;][]{1981ApJ...246..751K}. Thus only the brightest and/or most compact SS\,433-like systems will be detectable in galaxies like M\,82. 

\par
Strong blue- and redshifted optical emission lines like those seen in SS\,433 could potentially be used a diagnostic tool to verify the nature of our source. SS\,433 has an R band magnitude of 12 mag; the H\,$\alpha$ emission line is 10 times brighter than the continuum and 5\,000 km\,s$^{-1}$ wide. At the distance of M\,82, assuming similar intrinsic luminosity and extinction, the continuum of our source would not be detected with ground based instruments, but the H\,$\alpha$ line should be. Given that the radio luminosity of the M\,82 source is many times greater than that of SS\,433, it would not be surprising if the M82 source were also more luminous optically, increasing the chances of detecting the emission lines. However, if the M\,82 source suffers from more extinction there may also several bright infrared emission lines (Paschen $\alpha$ and Brackett $\gamma$ to B15) that could be used to identify it instead \citep{1979ApJ...234L.135T}.

\section{Conclusions}
We have shown that the recently discovered transient in M\,82 is relatively faint in X-rays given its radio luminosity. The lack of variability in its radio light curve also make it unlikely that the source is a normal BH transient. The source could very well be the extragalactic analogue of SS\,433 or a younger version of S\,26.
\\

\textbf{ACKNOWLEDGMENTS}\\
TDJ acknowledges support from a Stobie-SALT studentship, funded jointly by the NRF of South Africa, the British Council and the University of Southampton. We thank Rob Beswick, Michael Coriat and Martin Hardcastle for useful discussions and Elena Gallo and Dan Calvelo for providing tabulated X-ray and radio data for Galactic sources. We thank the referee for highlighting areas in the original draft which needed clarification. The GBI was operated by the National Radio Astronomy Observatory for the U.S. Naval Observatory and the Naval Research laboratory during the time period of the radio data used in Fig.\,\ref{CygX3_SS433} observations. This research has made use of data obtained from the Chandra Data Archive and software provided by the Chandra X-ray Center (CXC) in the application packages CIAO and ChIPS.

\bibliographystyle{mn2e}
\bibliography{Ref_list}

\begin{thebibliography}{}

\bibitem[\protect\citeauthoryear{{Begelman}, {King} \& {Pringle}}{{Begelman}
  et~al.}{2006}]{2006MNRAS.370..399B}
{Begelman} M.~C.,  {King} A.~R.,    {Pringle} J.~E.,  2006, \mnras, 370, 399

\bibitem[\protect\citeauthoryear{{Beswick}, {Riley}, {Marti-Vidal}, {Pedlar},
  {Muxlow}, {McDonald}, {Wills}, {Fenech} \& {Argo}}{{Beswick}
  et~al.}{2006}]{2006MNRAS.369.1221B}
{Beswick} R.~J.,  {Riley} J.~D.,  {Marti-Vidal} I.,  {Pedlar} A.,  {Muxlow}
  T.~W.~B.,  {McDonald} A.~R.,  {Wills} K.~A.,  {Fenech} D.,    {Argo} M.~K.,
  2006, \mnras, 369, 1221

\bibitem[\protect\citeauthoryear{{Bland} \& {Tully}}{{Bland} \&
  {Tully}}{1988}]{1988Natur.334...43B}
{Bland} J.,  {Tully} B.,  1988, \nat, 334, 43

\bibitem[\protect\citeauthoryear{{Brocksopp}, {Fender}, {McCollough}, {Pooley},
  {Rupen}, {Hjellming}, {de la Force}, {Spencer}, {Muxlow}, {Garrington} \&
  {Trushkin}}{{Brocksopp} et~al.}{2002}]{2002MNRAS.331..765B}
{Brocksopp} C.,  {Fender} R.~P.,  {McCollough} M.,  {Pooley} G.~G.,  {Rupen}
  M.~P.,  {Hjellming} R.~M.,  {de la Force} C.~J.,  {Spencer} R.~E.,  {Muxlow}
  T.~W.~B.,  {Garrington} S.~T.,    {Trushkin} S.,  2002, \mnras, 331, 765

\bibitem[\protect\citeauthoryear{{Brunthaler}, {Menten}, {Reid}, {Henkel},
  {Bower} \& {Falcke}}{{Brunthaler} et~al.}{2009}]{2009ATel.2078....1B}
{Brunthaler} A.,  {Menten} K.~M.,  {Reid} M.~J.,  {Henkel} C.,  {Bower} G.~C.,
    {Falcke} H.,  2009, The Astronomer's Telegram, 2078, 1

\bibitem[\protect\citeauthoryear{{Calvelo}, {Fender}, {Russell}, {Gallo},
  {Corbel}, {Tzioumis}, {Bell}, {Lewis} \& {Maccarone}}{{Calvelo}
  et~al.}{2010}]{2010MNRAS.409..839C}
{Calvelo} D.~E.,  {Fender} R.~P.,  {Russell} D.~M.,  {Gallo} E.,  {Corbel} S.,
  {Tzioumis} A.~K.,  {Bell} M.~E.,  {Lewis} F.,    {Maccarone} T.~J.,  2010,
  \mnras, 409, 839

\bibitem[\protect\citeauthoryear{{Corbel}, {Nowak}, {Fender}, {Tzioumis} \&
  {Markoff}}{{Corbel} et~al.}{2003}]{2003A&A...400.1007C}
{Corbel} S.,  {Nowak} M.~A.,  {Fender} R.~P.,  {Tzioumis} A.~K.,    {Markoff}
  S.,  2003, \aap, 400, 1007

\bibitem[\protect\citeauthoryear{{Dickey} \& {Lockman}}{{Dickey} \&
  {Lockman}}{1990}]{1990ARA&A..28..215D}
{Dickey} J.~M.,  {Lockman} F.~J.,  1990, \araa, 28, 215

\bibitem[\protect\citeauthoryear{{Dubner}, {Holdaway}, {Goss} \&
  {Mirabel}}{{Dubner} et~al.}{1998}]{1998AJ....116.1842D}
{Dubner} G.~M.,  {Holdaway} M.,  {Goss} W.~M.,    {Mirabel} I.~F.,  1998, \aj,
  116, 1842

\bibitem[\protect\citeauthoryear{{Fabrika}}{{Fabrika}}{2004}]{2004ASPRv..12...%
.1F}
{Fabrika} S.,  2004, Astrophysics and Space Physics Reviews, 12, 1

\bibitem[\protect\citeauthoryear{{Falcke}, {K{\"o}rding} \& {Markoff}}{{Falcke}
  et~al.}{2004}]{2004A&A...414..895F}
{Falcke} H.,  {K{\"o}rding} E.,    {Markoff} S.,  2004, \aap, 414, 895

\bibitem[\protect\citeauthoryear{{Fender}, {Corbel}, {Tzioumis}, {Tingay},
  {Brocksopp} \& {Gallo}}{{Fender} et~al.}{2002}]{2002ATel..107....1F}
{Fender} R.,  {Corbel} S.,  {Tzioumis} T.,  {Tingay} S.,  {Brocksopp} C.,
  {Gallo} E.,  2002, The Astronomer's Telegram, 107, 1

\bibitem[\protect\citeauthoryear{{Fender}, {Belloni} \& {Gallo}}{{Fender}
  et~al.}{2004}]{2004MNRAS.355.1105F}
{Fender} R.~P.,  {Belloni} T.~M.,    {Gallo} E.,  2004, \mnras, 355, 1105

\bibitem[\protect\citeauthoryear{{Fenech}, {Muxlow}, {Beswick}, {Pedlar} \&
  {Argo}}{{Fenech} et~al.}{2008}]{2008MNRAS.391.1384F}
{Fenech} D.~M.,  {Muxlow} T.~W.~B.,  {Beswick} R.~J.,  {Pedlar} A.,    {Argo}
  M.~K.,  2008, \mnras, 391, 1384

\bibitem[\protect\citeauthoryear{{Fiedler}, {Johnston}, {Spencer}, {Waltman},
  {Florkowski}, {Matsakis}, {Josties}, {Angerhofer}, {Klepczynski} \&
  {McCarthy}}{{Fiedler} et~al.}{1987}]{1987AJ.....94.1244F}
{Fiedler} R.~L.,  {Johnston} K.~J.,  {Spencer} J.~H.,  {Waltman} E.~B.,
  {Florkowski} S.~R.,  {Matsakis} D.~N.,  {Josties} F.~J.,  {Angerhofer} P.~E.,
   {Klepczynski} W.~J.,    {McCarthy} D.~D.,  1987, \aj, 94, 1244

\bibitem[\protect\citeauthoryear{{Fraser}, {Smartt}, {Crockett}, {Mattila},
  {Stephens} \& {Roth}}{{Fraser} et~al.}{2009}]{2009ATel.2131....1F}
{Fraser} M.,  {Smartt} S.~J.,  {Crockett} M.,  {Mattila} S.,  {Stephens}
  A.~G.-Y.~A.,    {Roth} K.,  2009, The Astronomer's Telegram, 2131, 1

\bibitem[\protect\citeauthoryear{{Freedman}, {Hughes}, {Madore}, {Mould},
  {Lee}, {Stetson}, {Kennicutt}, {Turner}, {Ferrarese}, {Ford}, {Graham},
  {Hill}, {Hoessel}, {Huchra} \& {Illingworth}}{{Freedman}
  et~al.}{1994}]{1994ApJ...427..628F}
{Freedman} W.~L.,  {Hughes} S.~M.,  {Madore} B.~F.,  {Mould} J.~R.,  {Lee}
  M.~G.,  {Stetson} P.,  {Kennicutt} R.~C.,  {Turner} A.,  {Ferrarese} L.,
  {Ford} H.,  {Graham} J.~A.,  {Hill} R.,  {Hoessel} J.~G.,  {Huchra} J.,
  {Illingworth} G.~D.,  1994, \apj, 427, 628

\bibitem[\protect\citeauthoryear{{Gallo}, {Fender}, {Miller-Jones}, {Merloni},
  {Jonker}, {Heinz}, {Maccarone} \& {van der Klis}}{{Gallo}
  et~al.}{2006}]{2006MNRAS.370.1351G}
{Gallo} E.,  {Fender} R.~P.,  {Miller-Jones} J.~C.~A.,  {Merloni} A.,  {Jonker}
  P.~G.,  {Heinz} S.,  {Maccarone} T.~J.,    {van der Klis} M.,  2006, \mnras,
  370, 1351

\bibitem[\protect\citeauthoryear{{Gallo}, {Fender} \& {Pooley}}{{Gallo}
  et~al.}{2003}]{2003MNRAS.344...60G}
{Gallo} E.,  {Fender} R.~P.,    {Pooley} G.~G.,  2003, \mnras, 344, 60

\bibitem[\protect\citeauthoryear{{Gehrels}}{{Gehrels}}{1986}]{1986ApJ...303..3%
36G}
{Gehrels} N.,  1986, \apj, 303, 336

\bibitem[\protect\citeauthoryear{{Hannikainen}, {Wu}, {Campbell-Wilson},
  {Hunstead}, {Lovell}, {McIntyre}, {Reynolds}, {Soria} \&
  {Tzioumis}}{{Hannikainen} et~al.}{2001}]{2001ESASP.459..291H}
{Hannikainen} D.,  {Wu} K.,  {Campbell-Wilson} D.,  {Hunstead} R.,  {Lovell}
  J.,  {McIntyre} V.,  {Reynolds} J.,  {Soria} R.,    {Tzioumis} T.,  2001, in
  {A.~Gimenez, V.~Reglero, \& C.~Winkler} ed., Exploring the Gamma-Ray Universe
  Vol.~459 of ESA Special Publication, {Radio emission from the X-ray transient
  XTE J1550-564}.
pp 291--294

\bibitem[\protect\citeauthoryear{{Hillwig} \& {Gies}}{{Hillwig} \&
  {Gies}}{2008}]{2008ApJ...676L..37H}
{Hillwig} T.~C.,  {Gies} D.~R.,  2008, \apjl, 676, L37

\bibitem[\protect\citeauthoryear{{King}, {Taam} \& {Begelman}}{{King}
  et~al.}{2000}]{2000ApJ...530L..25K}
{King} A.~R.,  {Taam} R.~E.,    {Begelman} M.~C.,  2000, \apjl, 530, L25

\bibitem[\protect\citeauthoryear{{Kong} \& {Chiang}}{{Kong} \&
  {Chiang}}{2009}]{2009ATel.2080....1K}
{Kong} A.~K.~H.,  {Chiang} Y.-K.,  2009, The Astronomer's Telegram, 2080, 1

\bibitem[\protect\citeauthoryear{{Kronberg}, {Biermann} \& {Schwab}}{{Kronberg}
  et~al.}{1981}]{1981ApJ...246..751K}
{Kronberg} P.~P.,  {Biermann} P.,    {Schwab} F.~R.,  1981, \apj, 246, 751

\bibitem[\protect\citeauthoryear{{Kronberg} \& {Sramek}}{{Kronberg} \&
  {Sramek}}{1985}]{1985Sci...227...28K}
{Kronberg} P.~P.,  {Sramek} R.~A.,  1985, Science, 227, 28

\bibitem[\protect\citeauthoryear{{Margon}}{{Margon}}{1984}]{1984ARA&A..22..507%
M}
{Margon} B.,  1984, \araa, 22, 507

\bibitem[\protect\citeauthoryear{{Margon}, {Ford}, {Katz}, {Kwitter}, {Ulrich},
  {Stone} \& {Klemola}}{{Margon} et~al.}{1979}]{1979ApJ...230L..41M}
{Margon} B.,  {Ford} H.~C.,  {Katz} J.~I.,  {Kwitter} K.~B.,  {Ulrich} R.~K.,
  {Stone} R.~P.~S.,    {Klemola} A.,  1979, \apjl, 230, L41

\bibitem[\protect\citeauthoryear{{McDonald}, {Muxlow}, {Wills}, {Pedlar} \&
  {Beswick}}{{McDonald} et~al.}{2002}]{2002MNRAS.334..912M}
{McDonald} A.~R.,  {Muxlow} T.~W.~B.,  {Wills} K.~A.,  {Pedlar} A.,
  {Beswick} R.~J.,  2002, \mnras, 334, 912

\bibitem[\protect\citeauthoryear{{Merloni}, {Heinz} \& {di Matteo}}{{Merloni}
  et~al.}{2003}]{Merloni03}
{Merloni} A.,  {Heinz} S.,    {di Matteo} T.,  2003, \mnras, 345, 1057

\bibitem[\protect\citeauthoryear{{Mitsuda}, {Inoue}, {Nakamura} \&
  {Tanaka}}{{Mitsuda} et~al.}{1989}]{1989PASJ...41...97M}
{Mitsuda} K.,  {Inoue} H.,  {Nakamura} N.,    {Tanaka} Y.,  1989, \pasj, 41, 97

\bibitem[\protect\citeauthoryear{{Muxlow}, {Beswick}, {Garrington}, {Pedlar},
  {Fenech}, {Argo}, {van Eymeren}, {Ward}, {Zezas} \& {Brunthaler}}{{Muxlow}
  et~al.}{2010}]{2010MNRAS.404L.109M}
{Muxlow} T.~W.~B.,  {Beswick} R.~J.,  {Garrington} S.~T.,  {Pedlar} A.,
  {Fenech} D.~M.,  {Argo} M.~K.,  {van Eymeren} J.,  {Ward} M.,  {Zezas} A.,
  {Brunthaler} A.,  2010, \mnras, 404, L109

\bibitem[\protect\citeauthoryear{{Muxlow}, {Beswick}, {Pedlar}, {Fenech},
  {Argo}, {Ward} \& {Zezas}}{{Muxlow} et~al.}{2009}]{2009ATel.2073....1M}
{Muxlow} T.~W.~B.,  {Beswick} R.~J.,  {Pedlar} A.,  {Fenech} D.,  {Argo} M.~K.,
   {Ward} M.~J.,    {Zezas} A.,  2009, The Astronomer's Telegram, 2073, 1

\bibitem[\protect\citeauthoryear{{Muxlow}, {Pedlar}, {Wilkinson}, {Axon},
  {Sanders} \& {de Bruyn}}{{Muxlow} et~al.}{1994}]{1994MNRAS.266..455M}
{Muxlow} T.~W.~B.,  {Pedlar} A.,  {Wilkinson} P.~N.,  {Axon} D.~J.,  {Sanders}
  E.~M.,    {de Bruyn} A.~G.,  1994, \mnras, 266, 455

\bibitem[\protect\citeauthoryear{{Pakull}, {Soria} \& {Motch}}{{Pakull}
  et~al.}{2010}]{2010Natur.466..209P}
{Pakull} M.~W.,  {Soria} R.,    {Motch} C.,  2010, \nat, 466, 209

\bibitem[\protect\citeauthoryear{{Predehl}, {Burwitz}, {Paerels} \&
  {Tr{\"u}mper}}{{Predehl} et~al.}{2000}]{2000A&A...357L..25P}
{Predehl} P.,  {Burwitz} V.,  {Paerels} F.,    {Tr{\"u}mper} J.,  2000, \aap,
  357, L25

\bibitem[\protect\citeauthoryear{{Romero}, {Torres}, {Kaufman Bernad{\'o}} \&
  {Mirabel}}{{Romero} et~al.}{2003}]{2003A&A...410L...1R}
{Romero} G.~E.,  {Torres} D.~F.,  {Kaufman Bernad{\'o}} M.~M.,    {Mirabel}
  I.~F.,  2003, \aap, 410, L1

\bibitem[\protect\citeauthoryear{{Soria}, {Pakull}, {Broderick}, {Corbel} \&
  {Motch}}{{Soria} et~al.}{2010}]{2010MNRAS.409..541S}
{Soria} R.,  {Pakull} M.~W.,  {Broderick} J.~W.,  {Corbel} S.,    {Motch} C.,
  2010, \mnras, 409, 541

\bibitem[\protect\citeauthoryear{{Stevens}, {Hannikainen}, {Wu}, {Hunstead} \&
  {McKay}}{{Stevens} et~al.}{2003}]{2003MNRAS.342..623S}
{Stevens} J.~A.,  {Hannikainen} D.~C.,  {Wu} K.,  {Hunstead} R.~W.,    {McKay}
  D.~J.,  2003, \mnras, 342, 623

\bibitem[\protect\citeauthoryear{{Strickland}, {Ponman} \&
  {Stevens}}{{Strickland} et~al.}{1997}]{1997A&A...320..378S}
{Strickland} D.~K.,  {Ponman} T.~J.,    {Stevens} I.~R.,  1997, \aap, 320, 378

\bibitem[\protect\citeauthoryear{{Telesco} \& {Harper}}{{Telesco} \&
  {Harper}}{1980}]{1980ApJ...235..392T}
{Telesco} C.~M.,  {Harper} D.~A.,  1980, \apj, 235, 392

\bibitem[\protect\citeauthoryear{{Thompson}, {Rieke}, {Tokunaga} \&
  {Lebofsky}}{{Thompson} et~al.}{1979}]{1979ApJ...234L.135T}
{Thompson} R.~I.,  {Rieke} G.~H.,  {Tokunaga} A.~T.,    {Lebofsky} M.~J.,
  1979, \apjl, 234, L135

\bibitem[\protect\citeauthoryear{{Walter}, {Weiss} \& {Scoville}}{{Walter}
  et~al.}{2002}]{2002ApJ...580L..21W}
{Walter} F.,  {Weiss} A.,    {Scoville} N.,  2002, \apjl, 580, L21

\bibitem[\protect\citeauthoryear{{Waltman}, {Ghigo}, {Johnston}, {Foster},
  {Fiedler} \& {Spencer}}{{Waltman} et~al.}{1995}]{1995AJ....110..290W}
{Waltman} E.~B.,  {Ghigo} F.~D.,  {Johnston} K.~J.,  {Foster} R.~S.,  {Fiedler}
  R.~L.,    {Spencer} J.~H.,  1995, \aj, 110, 290

\bibitem[\protect\citeauthoryear{{Wei{\ss}}, {Neininger}, {H{\"u}ttemeister} \&
  {Klein}}{{Wei{\ss}} et~al.}{2001}]{2001A&A...365..571W}
{Wei{\ss}} A.,  {Neininger} N.,  {H{\"u}ttemeister} S.,    {Klein} U.,  2001,
  \aap, 365, 571

\end{thebibliography}

\bsp

\label{lastpage}

\end{document}